\documentclass[preprint,12pt,sort&compress]{elsarticle}
\graphicspath{{./Fig/}}
\usepackage{amssymb}
\usepackage{amsthm}
\usepackage{amsfonts}
\usepackage{amsmath,bm} 
\usepackage{gensymb}	
\usepackage{graphicx}
\graphicspath{{./Fig/}}
\usepackage{dcolumn}
\usepackage{booktabs}

\usepackage{lmodern}           
\usepackage[T1]{fontenc}       
\usepackage[cp1250]{inputenc}  
\usepackage{verbatim} 	       
\usepackage{setspace}
\usepackage[colorlinks=true,
pdfstartview=FitV, linkcolor=cyan,
citecolor=cyan, urlcolor=cyan]{hyperref}

\journal{International Jounral of Engineering Sciences}

\usepackage[demo]{adjustbox}
\usepackage{xcolor}
\usepackage{atbegshi}
\AtBeginDocument{\AtBeginShipoutNext{\AtBeginShipoutDiscard}}	
\PassOptionsToPackage{hyphens}{url}\usepackage{hyperref} 

\begin{document}
	\pagestyle{empty} 
	\begin{titlepage}
		\color[rgb]{.4,.4,1}
		\hspace{5mm}

		\bigskip
		
		\hspace{15mm}
		\begin{minipage}{10mm}
			\color[rgb]{.7,.7,1}
			\rule{1pt}{226mm}
		\end{minipage}
		\begin{minipage}{133mm}
			\vspace{10mm}        
			\color{black}
			\sffamily
			\LARGE\bfseries Stress-Driven Modelling  \\[-0.3\baselineskip]  of Nonlocal Thermoelastic  \\[-0.3\baselineskip]  Behavior of Nanobeams
			
			\vspace{5mm}
			{\large {Preprint of the article published in \\[-0.4\baselineskip] International Jounral of Engineering Sciences \\[-0.1\baselineskip] 126, May 2018, 53-67 }} 
			
			\vspace{10mm}        
			{\large Raffaele Barretta, \\[-0.4\baselineskip] \textsc{Marko \v{C}ana\dj{}ija}, \\[-0.4\baselineskip] Raimondo Luciano, \\[-0.4\baselineskip] Francesco Marotti de Sciarra} 
			
			\large
			
			\vspace{40mm}
			\vspace{5mm}
			
			\small
			\url{https://doi.org/10.1016/j.ijengsci.2018.02.012}
			
			\textcircled{c} 2018. This manuscript version is made available under the CC-BY-NC-ND 4.0 license \url{http://creativecommons.org/licenses/by-nc-nd/4.0/}
			\hspace{30mm} 
			\color[rgb]{.4,.4,1} 
		\end{minipage}
	\end{titlepage}

\begin{frontmatter}


\author[UN]{Raffaele Barretta}
\author[URI]{Marko \v{C}ana\dj{}ija \corref{cor}}
\ead{marko.canadija@riteh.hr}
\cortext[cor]{Corresponding author. Tel.: +385-51-651-496; Fax.: +385-51-651-490}
\author[LZ]{Raimondo Luciano}
\author[UN]{Francesco Marotti de Sciarra}

\address[UN]{Department of Structures for Engineering and Architecture,University of Naples Federico II,Via Claudio 21,80121 Naples,Italy}
\address[URI]{Faculty of Engineering,Department of Engineering Mechanics,University of Rijeka,Vukovarska 58,51000 Rijeka,Croatia}
\address[LZ]{Department of Civil and Mechanical Engineering, University of Cassino and Southern Lazio, via G. Di Biasio 43, 03043 Cassino (FR), Italy}

\begin{abstract}
A consistent stress-driven nonlocal integral model for nonisothermal structural analysis of elastic nano- and microbeams is proposed. Most nonlocal models of literature are strain-driven and it was shown that such approaches can lead toward a number of difficulties. Following recent contributions within the isothermal setting, the developed model abandons the classical strain-driven methodology in favour of the modern stress-driven elasticity theory by G. Romano and R. Barretta. This effectively circumvents issues associated with strain-driven formulations. The new thermoelastic nonlocal integral model is proven to be equivalent to an adequate set of differential equations, accompanied by higher-order constitutive boundary conditions, when the special Helmholtz averaging kernel is adopted in the convolution. The example section provides several applications, thus enabling insight into performance of the formulation. Exact nonlocal solutions are established, detecting also new benchmarks for thermoelastic numerical analyses. 
\end{abstract}

\end{frontmatter}

\section{Introduction}
Research breakthroughs in nanotechnology over recent years have caused an increased interest in the mechanical behaviour of structures at nanoscale as well. Most commonly, sensors that measure forces or displacements are analysed. However, the mechanics of such structures fails outside usual macroscale mechanical principles. Origins of such behaviour are manifold. The structures at the nanoscale have discrete nature manifested in the form of atoms and interactions between atoms so that continuum mechanics can have limited success in describing discrete nanostructures. Furthermore, forces that are completely irrelevant at the macroscale, van der Waals forces for example, can dominate the nanoscale behaviour. As a consequence, size effects start to appear. The research community has been very active lately trying to capture this behaviour by accounting for the nonlocal nature of the phenomenon. Although such a claim can be made for the problems involving isothermal deformation processes, when the nonisothermal problems are concerned the results are not so numerous. For a short review of the existing nonisothermal models, see \cite{Canadija16b} for statical and \cite{Zenkour14c} for dynamical problems.

The existing methods are almost exclusively based on gradient methods and are applied to a variety of different problems, see \cite{Barretta16,Sedighi2014,Simsek13,Papargyri03,Romanoff16,Sedighi14} for a start. As a cornerstone of most approaches, the assumption that the strain gradient also contributes to the stress state in a point is utilized \cite{Sciarra14}. To tune-up theoretical models to experimentally observed behaviour, the existence of the so-called nonlocal (small-size) constitutive parameter is postulated.  However, the literature survey in \cite{Canadija17} shows surprisingly low number of results on determination of the nonlocal parameter obtained either by means of an experimental or a theoretical procedure. The latter paper also shows that nonlocal behaviour of the nanostructures do exists even in the case of harmonic interatomic potential and may be attributed to the discrete nature of the structure. Moreover, it demonstrates that the gradient methods have difficulties matching simulated bending of carbon nanotubes.

A gradient method, widely adopted to describe size-dependent phenomena 
in nanostructures, is based on Eringen differential model (EDM) 
associated with the strain-driven nonlocal integral theory conceived in \cite{Eringen83}.
As shown more than a decade ago \cite{Peddieson03}, although nanosensors are usually designed as a cantilever nanobeam with a tip force and nonlocal effects are readily experimentally observed,  EDM is not adequate to assess size effects.
Elastic responses associated with EDM are technically unacceptable, 
as discussed in \cite{Challamel08,Challamel16,Fernandez16,Eptaimeros16,Koutsoumaris16}. 
Strain-driven nonlocal integral theory, introduced and successfully adopted by Eringen 
to study (in unbounded domains) screw dislocations and surface waves, is inapplicable to Structural Mechanics
\cite{Romano16}. 

This conclusion is due to the fact that the elastostatic problem of a 
continuous structure defined on a bounded domain,
formulated by the strain-driven nonlocal integral model, admits no solution for all 
static schemes of engineering interest.
Nonlocal (strain-driven) stress fields in bounded structural domains
are indeed not included in the affine manifold of equilibrated stresses fields,
as recognized in literature 
\cite{VilaIJSS2017,BaratiIJESci2017,FathiBraz2017,SourkiEPJP2017,ZhuIJESci2017,FernIJESci2017, MorassiMSSP, WangIJESci2017, ZhangIJMSci2017, XuIJESci2017, LiLiThinWall2018}
on the basis of the original contribution in \cite{Romano17}. 
All difficulties can be overcome by resorting to the innovative stress-driven
nonlocal integral theory recently proposed by G. Romano and R. Barretta
\cite{Romano2017c}.
According to the stress-driven approach, the nonlocal elastic strain field
is the convolution between the stress field and a suitable averaging kernel.
Properties and merits of the stress-driven strategy in comparison with
strain-driven formulations can be found in \cite{Romano17b,RomanoBarDia2017}. 
Transverse free vibrations of Bernoulli-Euler nanobeams are investigated
in \cite{ApuzzoBar17} by stress-driven integral approach.

The research at hand aims to provide a well-posed nonlocal integral formulation of nanobeam mechanics in nonisothermal regime. 
It is assumed that the nanobeam can be described by Bernoulli-Euler kinematics. 
This kind of model has not been previously addressed in literature. The method presents an extension of the contributions presented in \cite{Romano2017c} 
and therefore can be categorized as a stress-driven model.

\section{Kinematics of Nonisothermal Bernoulli-Euler Beams}
\label{sec:Kinem}
\begin{figure}
	\centering
	\includegraphics[scale=0.4]{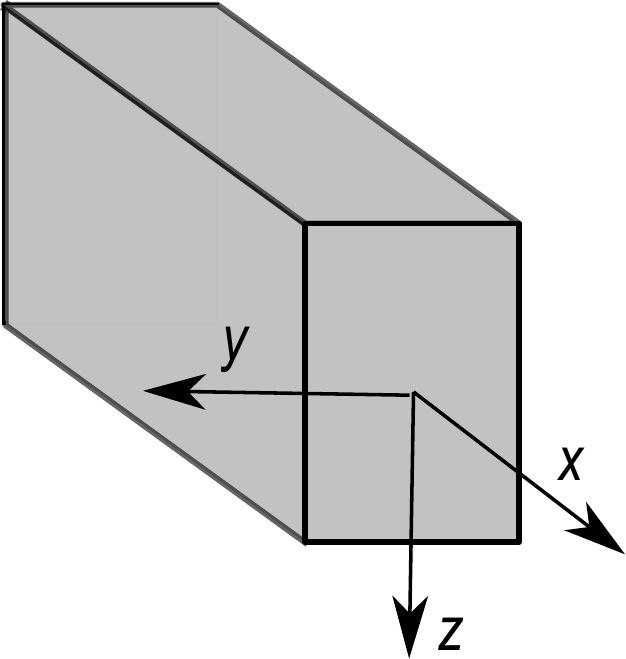}
	\caption{Coordinate system of a Bernoulli-Euler nanobeam} 
	\label{fig:Beam_CS}
\end{figure}

This section will introduce the notation and provide well-known governing equations that will serve as a starting point for the nonlocal integral formulation introduced in \textsection\ref{sec_nonlocal}. In the subsequent analysis, a straight nanobeam made of the material with the coefficient of thermal expansion $\alpha$ is considered. The nanobeam's cross-section $\Omega$ is assumed to lay in the $y-z$ plane while the longitudinal axis is denoted with $x$, Fig.~\ref{fig:Beam_CS}. The longitudinal axis $x$ is assumed to pass through centroids of the cross-section.  Due to the Bernoulli-Euler assumption, only normal stresses directed along $x$ axis exist and these will be denoted by $\sigma$. To simplify elaborations and not to lose focus on the central problem, bending in $x-z$ plane will be considered. Finally, the axes $y$ and $z$ are assumed to be the principal axes of inertia of the cross-section, thus giving $I_\mathrm{yz}=0$. As usual, $A$ is the area and $I_\mathrm{y}$ is the second moment of area of the cross-section, respectively.

The displacement $\mathbf{u}=\left\lbrace u_x, u_y, u_z\right\rbrace $ of any point of the beam can be evaluated as:
\begin{equation}\label{eq:DispField}
u_x(x,z)=u(x,z)=u_0(x)+\varphi(x) z  , \quad u_z(x)=w(x), \quad u_y=0,
\end{equation}
where $\varphi(x)$ denotes the angle of rotation of the cross-section. As the above equation indicates, the longitudinal displacement $u$ is composed of two parts. The first part $u_0$ represents the average displacement of the cross section, defined as:
\begin{equation}\label{eq:Avgu}
u_0(x)=\frac{1}{A}\int_\Omega u(x,z) \mathrm{d}A.
\end{equation}
When introduced into the Eq.(\ref{eq:Avgu}), the second part of displacement field $\varphi(x) z$ in Eq.~(\ref{eq:DispField}) gives $\varphi(x) \int_\Omega  z \mathrm{d}A = \varphi(x) \; S_z$. If the origin of the axis $z$ is positioned on the symmetry line passing through the centroid of the cross-section, the first moment of area $S_z$ vanishes providing the result in Eq.~(\ref{eq:Avgu}). 

Due to the Bernoulli-Euler assumption, the cross-section remains plane. This effectively enforces vanishing shear strains, thus providing a link between the derivatives of the longitudinal and the transversal displacement as:
\begin{equation}\label{eq:PhiDispRelation}
0=\gamma _\mathrm{xz} (x)=\partial _x w(x)+\partial _z u(x,z)=w^{(1)}(x) + \varphi(x),
\end{equation}
where Eq.(\ref{eq:DispField})$_1$ was utilized. The apex $^{(n)}$ denotes the $n$-th derivative with respect to the longitudinal coordinate $x$. With the link between the transversal displacement and the rotation established, the strains in a point can be now conveniently obtained by differentiating the axial displacement field with respect to the longitudinal coordinate as:
\begin{equation}\label{eq:Epsilon}
\varepsilon (x,z)=\partial _x u(x,z)=u_0^{(1)}(x) +\varphi ^{(1)}(x)z=u_0^{(1)}(x) - w ^{(2)}(x)z. \\
\end{equation}
The above introduced kinematic framework can account for both isothermal and nonisothermal class of problems.  

Note that the normal strain can be additively separated into an axial and a bending part. The axial part of the normal strain can be naturally introduced as the centroidal normal strain by setting $z=0$ in Eq.~(\ref{eq:Epsilon}), providing:
\begin{equation}\label{eq:Eps0}
\varepsilon_0 =  u_0^{(1)}.
\end{equation}
Likewise, a standard procedure shows that the curvature $\chi$ of the deformed axis of the beam is related to the transverse displacement $w$ as:
\begin{equation}\label{eq:EpsBend}
\quad w^{(2)}= \chi.
\end{equation}
This allows definition of the bending part of the normal strain as $z  \; \chi(x)$. Hence, the normal strain can be represented as:
\begin{equation}\label{eq:AxBend}
\varepsilon (x,z)=\varepsilon_0(x) -z  \; \chi(x).
\end{equation}

\section{Stresses and Deflections in Local Nonisothermal Bernoulli-Euler Beams}\label{sec:local}
As this point, calculation of stresses in the local Bernoulli-Euler nonisothermal beam should be summarized. The case in which the beam is free from external mechanical loads, but subjected to the non-homogeneous temperature field variation $\Delta \theta (x,z)$ is considered. In such type of problems stresses can arise as well. The first step involves separation of the normal strain into a part that describes thermal dilatation due to free expansion (or contraction) of a beam $\varepsilon_{\mathrm{th}}(x,z)=\alpha \Delta \theta(x,z)$ and the one caused by the elastic effects $\varepsilon_{\mathrm{el}}(x,z)=\sigma(x,z)/E$:
\begin{equation}\label{eq:StrainAddSep}
\varepsilon=\varepsilon_{\mathrm{th}}+\varepsilon_{\mathrm{el}}.
\end{equation}
It should be emphasized that only elastic (mechanical) part of the strain is used to calculate stresses; free thermal elongation does not contribute to stress evolution. Having in mind kinematics of the Bernoulli-Euler beam as described in \textsection\ref{sec:Kinem}, the sum of thermal and elastic strain should equal axial and bending deformation, Eq.~(\ref{eq:AxBend}):
\begin{equation}\label{eq:StrainLocEq}
\alpha \Delta \theta(x,z)+ \frac{\sigma(x,z)}{E}=\varepsilon_0(x)-z \; \chi(x),
\end{equation}
so the normal stress is obtained as $\sigma(x,z)= -\alpha E \Delta \theta(x,z) + \varepsilon_0(x) E -z E \; \chi(x)$. In the absence of external mechanical loading and upon introduction of the normal stress, equilibrium equations provide the axial strain and the curvature:
\begin{equation}\label{eq:LocEq}
\begin{array}{l}
\displaystyle\int_\Omega \sigma dA = 0 \quad \rightarrow \quad  \varepsilon_0  =  \frac{1}{A E}\displaystyle\int_\Omega \alpha E \Delta \theta dA , 
\vspace{6pt}\cr
\displaystyle\int_\Omega \sigma z dA = 0 \quad \rightarrow \quad \chi =  - \frac{1}{I_\mathrm{y} E} \displaystyle\int_\Omega \alpha E \Delta \theta z dA.
\end{array}
\end{equation}
With known $\varepsilon_0$ and $\chi$, the normal stress is rewritten as:
\begin{equation}\label{eq:IntForces2T}
\sigma=- \alpha \Delta \theta E +\frac{\displaystyle\int_{\Omega} \alpha \Delta \theta E\mathrm{d}A}{A} + \frac{\displaystyle\int_{\Omega} \alpha E \Delta \theta z \mathrm{d}A}{I_\mathrm{y}}z,
\end{equation}
see classical textbooks on the subject \cite{Boley1967,Noda03} for the in-depth discussion. On the other side, if the beam is subjected to simultaneous action of mechanical and thermal loads, the stresses are obtained by the simple superposition procedure:
\begin{equation}\label{eq:IntForcesMT}
\sigma=- \alpha \Delta \theta E +\frac{N+\displaystyle\int_{\Omega} \alpha \Delta \theta E\mathrm{d}A}{A} + \frac{M+\displaystyle\int_{\Omega} \alpha E \Delta \theta z \mathrm{d}A}{I_\mathrm{y}}z,
\end{equation}
where $N$ is the the axial force and $M$ is the bending moment in the corresponding cross-section. The two integrals $\int_{\Omega} \alpha \Delta \theta E\mathrm{d}A$ and $\int_{\Omega} \alpha E \Delta \theta z \mathrm{d}A$ are usually named as the thermal axial force $N_\mathrm{th}$ and the thermal bending moment $M_\mathrm{th}$.

The deflection line can be obtained using Eq.~(\ref{eq:EpsBend}) and results Eq.~(\ref{eq:LocEq})$_2$, thus providing the differential equation:
\begin{equation}\label{eq:DefLine}
w^{(2)}=- \frac{M+\displaystyle\int_{\Omega} \alpha E \Delta \theta z \mathrm{d}A}{E I_\mathrm{y}},
\end{equation}
accompanied by a suitable set of boundary conditions.

\section{Stress-driven nonlocal thermoelasticity}\label{sec_nonlocal}

The framework described so far should be now extended to account for nonlocality present at the nanoscale level. The first step again involves separation of the strain into a part that describes dilatation due to free expansion (or contraction) of a beam $\varepsilon_{\mathrm{th}}(x,z)=\alpha \Delta \theta(x,z)$ and the one caused by the stresses $\varepsilon_{\mathrm{el}}$, Eq.~(\ref{eq:StrainAddSep}). In contrast to the earlier works on the nonlocal beam mechanics cited in the Introduction, the present approach is stress-driven. Accordingly, the elastic strains originating from stresses are assumed to be of the nonlocal nature and defined as a convolution \cite{Romano17b}:
\begin{equation}\label{eq:NonlocalStrain}
\varepsilon_{\mathrm{el}}(x,z)= \int_{0}^{L} \phi ({x-\xi)}  E^{-1} \sigma(\xi,z) \mathrm{d} \xi ,
\end{equation}
where the kernel function $\phi ({x)}$ is
\begin{equation}\label{eq:Kernel}
\phi (x) = \frac{1}{2 L_\lambda} \exp(-\frac{\left| x\right| }{ L_\lambda }).
\end{equation}
The characteristic length is defined as $L_\lambda= \lambda L>0$, i.e. as the product of the nonlocal parameter $\lambda>0$ and the beam's length $L$. Due to the exponential nature of the kernel function, the nonlocal effects in the vicinity of the corresponding point have significantly more influence on the strain than points situated at larger distance. Note that above assumption accounts only for nonlocality in the longitudinal direction. Variation of stresses in the transverse directions does not contribute to the nonlocal normal strain in the present formulation.

Since the decomposition Eq.~(\ref{eq:AxBend}) remains valid, the normal strain provides the equality:
\begin{equation}\label{eq:StrainEq}
\alpha \Delta \theta(x,z)+\int_{0}^{L} \phi ({x-\xi)}  E^{-1} \sigma(\xi,z) \mathrm{d} \xi=\varepsilon_0(x)-z \; \chi(x),
\end{equation}
where the thermal strain and the nonlocal elastic strain definition were used. If the latter equation is integrated over the cross section, it follows:
\begin{equation}\label{eq:StrainEqIntA}
\int _\Omega \alpha \Delta \theta(x,z) \mathrm{d}A + \int _\Omega \int_{0}^{L} \phi ({x-\xi)}  E^{-1} \sigma(\xi,z) \mathrm{d} \xi \mathrm{d}A =\int _\Omega  \varepsilon_0(x) \mathrm{d}A - \int _\Omega z \; \chi(x)\mathrm{d}A.
\end{equation}
The proposition that the cross-section axes are central and principal axes imply that the second term on the right hand side vanishes. In the second term on the left hand side, the equilibrium equation $\int_\Omega \sigma \mathrm{d} A = N$ can be employed. This yields:
\begin{equation}\label{eq:Eps0a}
\varepsilon_0(x)= \frac{1}{A} \left\lbrace \int _\Omega \alpha \Delta \theta(x,z) \mathrm{d}A + \int_{0}^{L} \phi ({x-\xi)}  E^{-1} N(\xi) \mathrm{d} \xi  \right\rbrace  .
\end{equation}
In the same manner, multiplication of Eq.~(\ref{eq:StrainEq}) by the transverse coordinate $z$, the equilibrium equation $\int_\Omega \sigma z \mathrm{d} A = M$ and subsequent integration over the cross section will provide the curvature as:
\begin{equation}\label{eq:Curvature}
\chi(x)= - \frac{1}{I_\mathrm{y}} \left\lbrace \int _\Omega \alpha \Delta \theta(x,z) z \mathrm{d}A + \int_{0}^{L} \phi ({x-\xi)}  E^{-1} M(\xi) \mathrm{d} \xi  \right\rbrace  .
\end{equation}
Obviously, the equilibrium conditions are explicitly enforced in the formulation. In order to simplify notation, at this point we introduce the following functions:
\begin{equation}\label{eq:PhiNM}
\varPhi_\mathrm{N}(x)=  \frac{1}{A} \int _\Omega \alpha \Delta \theta(x,z) \mathrm{d}A, \quad \varPhi_\mathrm{M}(x)=  \frac{1}{I_\mathrm{y}}\int _\Omega \alpha \Delta \theta(x,z) z \mathrm{d}A.
\end{equation}
Now, accounting for Eq.~(\ref{eq:Eps0}, \ref{eq:EpsBend}), a decoupled system of ordinary differential equations governing nonlocal mechanical behaviour of the nanobeam is obtained:
\begin{equation}\label{eq:ODE}
\begin{array}{l}
\displaystyle
u_0^{(1)}-\varPhi_\mathrm{N}= \int_{0}^{L} \phi ({x-\xi)}  \frac{N(\xi) }{EA} \mathrm{d} \xi , 
\vspace{6pt}\cr
\displaystyle
-w^{(2)}-\varPhi_\mathrm{M} = \int_{0}^{L} \phi ({x-\xi)}  \frac{M(\xi)}{EI_\mathrm{y}} \mathrm{d} \xi .
\end{array}
\end{equation}
The solutions of the above integrals are provided in \cite{Polyanin98}. For the average axial displacement this is:
\begin{equation}\label{eq:Solutionu0}
 \frac{N}{L_\lambda^{2} E A}  = \varPhi_\mathrm{N} ^{(2)} -  u_0^{(3)}  + \frac{1}{L_\lambda^{2}}\left(u_0^{(1)}-\varPhi_\mathrm{N}\right).
\end{equation}
This differential equation must be augmented with the constitutive boundary conditions:
\begin{equation}\label{eq:BCu0}
\begin{array}{l}
\displaystyle
u_0^{(2)}(0)-\varPhi_\mathrm{N}^{(1)}(0) -\frac{1}{L_\lambda}\left( u_0^{(1)}(0)-\varPhi_\mathrm{N}(0) \right) =0, 
\vspace{6pt}\cr
\displaystyle
u_0^{(2)}(L)-\varPhi_\mathrm{N}^{(1)}(L) +\frac{1}{L_\lambda}\left( u_0^{(1)}(L)-\varPhi_\mathrm{N}(L) \right) =0.
\end{array}
\end{equation}
For the proof of the above integral to differential transformation, see Remark 1 below for a detailed explanation. Even in the isothermal problems, the boundary conditions are of the constitutive character since they include a material property - the nonlocal parameter $L_\lambda$. In the nonisothermal problems, an additional material property is included in $\varPhi_\mathrm{N}$ and $\varPhi_\mathrm{M}$, the coefficient of heat expansion $\alpha$. 

The transverse displacements can be dealt with in the same manner:
\begin{equation}\label{eq:v}
\frac{M}{L_\lambda^{2} E I_\mathrm{y}} = w^{(4)}+\varPhi_\mathrm{M}^{(2)} + \frac{1}{L_\lambda^{2}}(- w^{(2)}-\varPhi_\mathrm{M} ),
\end{equation}
with the boundary conditions:
\begin{equation}\label{eq:BCv}
\begin{array}{l}
\displaystyle
-w^{(3)}(0)-\varPhi_\mathrm{M}^{(1)} (0) -\frac{1}{L_\lambda} \left(- w^{(2)}(0)-\varPhi_\mathrm{M}(0) \right) =0, 
\vspace{6pt}\cr
\displaystyle
-w^{(3)}(L)-\varPhi_\mathrm{M}^{(1)}(L) +\frac{1}{L_\lambda} \left( -w^{(2)}(L)-\varPhi_\mathrm{M}(L) \right) =0. \\
\end{array}
\end{equation}
If the problem is statically indeterminate, the set of differential equations Eqs.(\ref{eq:Solutionu0}, \ref{eq:v}) and boundary conditions Eqs.(\ref{eq:BCu0}, \ref{eq:BCv}) can be solved upon providing additional kinematic or static boundary conditions used in the standard beam theory.

\textbf{Remark 1.} The equivalence of Eq.~(\ref{eq:ODE})$_1$ and Eqs.~(\ref{eq:Solutionu0}, \ref{eq:BCu0}) is obtained by utilizing results \cite{Polyanin98}. However, the procedure in \cite{Polyanin98} demonstrates only necessity of existence of these conditions. To provide uniqueness of the solution, the proof in \cite{Romano2017c} in now generalized and applied to the problem at hand. In particular, consider an integral function:
\begin{equation}\label{eq:Rem1-1}
f(x)=\int_{x_1}^{x_2} \phi(x-y) g(y) \mathrm{d}y, \quad \mathrm{where}  \quad \phi (x-y) = \frac{1}{2 L_\lambda} \exp(-\frac{\left| x-y \right| }{ L_\lambda }),
\end{equation}
where $L_\lambda$ is a constant. In the first step, the above integral is additively separated into two parts: 
\begin{equation}\label{eq:Rem1-2}
\begin{array}{l}
f_1(x)=\displaystyle\int_{x_1}^{x} \phi(x-y) g(y) \mathrm{d}y=\displaystyle\int_{x_1}^{x} \frac{1}{2 L_\lambda} \exp(\frac{ y-x }{ L_\lambda }) g(y) \mathrm{d}y, 
\vspace{6pt}\cr
f_2(x)=\displaystyle\int_{x}^{x_2} \phi(x-y) g(y) \mathrm{d}y=\displaystyle\int_{x}^{x_2} \frac{1}{2 L_\lambda} \exp(\frac{ x-y }{ L_\lambda }) g(y) \mathrm{d}y, \\
\end{array}
\end{equation}
i.e., $f(x)=f_1(x)+f_2(x)$. In that way, the modulus in the kernel function $\phi (x-y)$ is removed. This implies following values of these functions at the boundary:
\begin{equation}\label{eq:Rem1-3}
\begin{array}{l}
f_2(x_1)=f(x_1), \quad f_1(x_1)=0,\\
f_1(x_2)=f(x_2), \quad f_2(x_2)=0.
\end{array}
\end{equation}
By virtue of the Leibniz integral rule, differentiation of Eq.~(\ref{eq:Rem1-2}) with respect to the longitudinal coordinate provides:
\begin{equation}\label{eq:Rem1-4}
\begin{array}{l}
\displaystyle
f_1'(x)= \frac{1}{2 L_\lambda} g(x) - \frac{1}{L_\lambda} \displaystyle\int_{x_1}^{x} \frac{1}{2 L_\lambda} \exp(\frac{ y-x }{ L_\lambda }) g(y) \mathrm{d}y = \frac{1}{2 L_\lambda} g(x) -\frac{1}{L_\lambda} f_1(x),
\vspace{6pt}\cr
\displaystyle
f_2'(x)= -\frac{1}{2 L_\lambda} g(x) + \frac{1}{L_\lambda} \displaystyle\int_{x}^{x_2} \frac{1}{2 L_\lambda} \exp(\frac{ x-y }{ L_\lambda }) g(y) \mathrm{d}y = -\frac{1}{2 L_\lambda} g(x) + \frac{1}{L_\lambda} f_2(x).  \\
\end{array}
\end{equation}
Thus, the first derivative $f'(x)$ is then:
\begin{equation}\label{eq:Rem1-5}
f'(x)=f_1'(x)+f_2'(x)=\frac{1}{L_\lambda} (f_2(x)-f_1(x)). 
\end{equation}
In the same manner, second derivatives are:
\begin{equation}\label{eq:Rem1-6}
\begin{array}{l}
\displaystyle
f_1''(x)= \frac{1}{2 L_\lambda} g'(x) - \frac{1}{2 L_\lambda^2} g(x) + \frac{1}{L_\lambda^2} \displaystyle\int_{x_1}^{x} \frac{1}{2 L_\lambda} \exp(\frac{ y-x }{ L_\lambda }) g(y) \mathrm{d}y 
\vspace{6pt}\cr
\displaystyle
f_2''(x)= -\frac{1}{2 L_\lambda} g'(x) - \frac{1}{2 L_\lambda^2} g(x) + \frac{1}{L_\lambda^2} \displaystyle\int_{x}^{x_2} \frac{1}{2 L_\lambda} \exp(\frac{ x-y }{ L_\lambda }) g(y) \mathrm{d}y.
\end{array}
\end{equation}
Now, $f''(x)=f_1''(x)+f_2''(x)$ gives:
\begin{equation}\label{eq:Rem1-7}
\frac{1}{L_\lambda^2}g(x)=\frac{1}{L_\lambda^2}f(x)-f''(x).
\end{equation}
The necessary boundary conditions are obtained from values of functions at the boundary, Eqs.~(\ref{eq:Rem1-3})  and the first derivative Eq.~(\ref{eq:Rem1-5}) as:
\begin{equation}\label{eq:Rem1-8}
f'(x_1)=\frac{1}{L_\lambda} f(x_1), \quad f'(x_2)= -\frac{1}{L_\lambda} f(x_2). 
\end{equation}
To prove the uniqueness of the solution, the solution of the homogeneous part of Eq.~(\ref{eq:Rem1-7}) is sought:
\begin{equation}\label{eq:Rem1-9}
f(x)-L_\lambda^2 f''(x)=0.
\end{equation}
The general form of the solution to the above differential equation is:
\begin{equation}\label{eq:Rem1-10}
f(x)=C_1+C_2 e^{x/L_\lambda^2}.
\end{equation}
Enforcement of the boundary conditions Eq.~(\ref{eq:Rem1-8}) results in:
\begin{equation}\label{eq:Rem1-11}
\begin{array}{l}
\displaystyle
\frac{1}{L_\lambda^2} C_2 e^{x_1/L_\lambda^2}=\frac{1}{L_\lambda} (C_1+C_2 e^{x_1/L_\lambda^2}),
\vspace{6pt}\cr
\displaystyle
\frac{1}{L_\lambda^2} C_2 e^{x_2/L_\lambda^2}=-\frac{1}{L_\lambda} (C_1+C_2 e^{x_2/L_\lambda^2}).
\end{array}
\end{equation}
Determination of constants $C_1$ and $C_2$ from the above system gives trivial solution. Now, if there are two solutions to the non-homogeneous problem Eq. ~(\ref{eq:Rem1-7}) and they are denoted as $F_1(X)$ and $F_2(x)$, then the difference of two solutions must give solution of the homogeneous problem:
\begin{equation}\label{eq:Rem1-12}
F_1(x)-F_2(x)=C_1+C_2 e^{x/L_\lambda^2},
\end{equation}
what upon enforcement of trivial solution ensures:
\begin{equation}\label{eq:Rem1-13}
F_1(x)=F_2(x),
\end{equation}
i.e. that the initial assumption about two different solutions was false and that only one solution exists. This proves uniqueness of the solution.


\textbf{Remark 2.} After determination of the displacement fields, calculation of stresses is straightforward. Namely,  Eq.~(\ref{eq:StrainEq}) can be transformed by introducing kinematic constraints Eq.~(\ref{eq:Eps0}, \ref{eq:EpsBend}) into:
\begin{equation}\label{eq:Stress1}
\int_{0}^{L} \phi ({x-\xi)}  E^{-1} \sigma(\xi) \mathrm{d} \xi=-\alpha \Delta \theta(x,z)+u_0^{(1)}(x)-z \; w^{(2)}.
\end{equation}
The stress field now follow as
\begin{equation}\label{eq:Stress2}
\sigma(x,z) = E\left\lbrace  -L_\lambda^{2} ((-\alpha \Delta \theta(x,z))^{(2)}+u_0^{(3)}(x)-z \; w^{(4)})  -\alpha \Delta \theta(x,z)+u_0^{(1)}(x)-z \; w^{(2)} \right\rbrace ,
\end{equation}
while the accompanying constitutive boundary conditions of type Eq.~(\ref{eq:BCu0}, \ref{eq:BCv}) were already accounted for in the evaluation of displacement fields.

In the absence of external forces, the thermal stresses can be alternatively evaluated from the convolution Eq.~(\ref{eq:StrainEq}), along with Eqs.~(\ref{eq:Eps0a}, \ref{eq:Curvature}) and definitions of Eq.~(\ref{eq:PhiNM}) by setting $N=0, M=0$:
\begin{equation}\label{eq:ThStress1}
\int_{0}^{L} \phi ({x-\xi)}  E^{-1} \sigma(\xi) \mathrm{d} \xi= -\alpha \Delta \theta(x,z)+\varPhi_\mathrm{N}(x)-z \; \varPhi_\mathrm{M}(x).
\end{equation}
Thermal stresses are now:
\begin{equation}\label{eq:ThStress2}
\sigma = E \left\lbrace -L_\lambda^{2} ((-\alpha \Delta \theta(x,z))^{(2)}+\varPhi_\mathrm{N}^{(2)}(x)-z \; \varPhi_\mathrm{M}^{(2)}(x))  -\alpha \Delta \theta(x,z)+\varPhi_\mathrm{N}(x)-z \; \varPhi_\mathrm{M}(x) \right\rbrace .
\end{equation}
It is easy to verify that if the product $\alpha \Delta \theta$ is constant, thermal stresses do not arise. The similar result follows in some other situations. For instance, when the temperature field is independent of the transverse coordinate $z$ and simultaneously linearly dependent on the longitudinal coordinate $x$ (vice-versa is also true).

\section{Examples}
\label{Examples}
To illustrate behaviour of the proposed methodology, several examples are considered. Introductory examples are simpler providing an interpretation of familiar topics from the standard courses of mechanics. Subsequently, more complex examples are given. The examples are solved by the aid of Wolfram Mathematica software.
\subsection{Simultaneous action of uniform temperature and axial force}
\label{Ex2}
A homogeneous beam of a symmetric cross-section of area $A$ and Young's modulus $E$ is subjected to a uniform temperature change $\Delta \theta$. At $x=0$ beam's expansion is constrained,  $u_0=0$. Otherwise the beam can expand freely. It is additionally loaded with the tensile longitudinal force $P$ at the beam's other end, $x=L$. It is necessary to determine the extension and stresses in the beam. 

\begin{figure}
	\centering
	\includegraphics[scale=0.50]{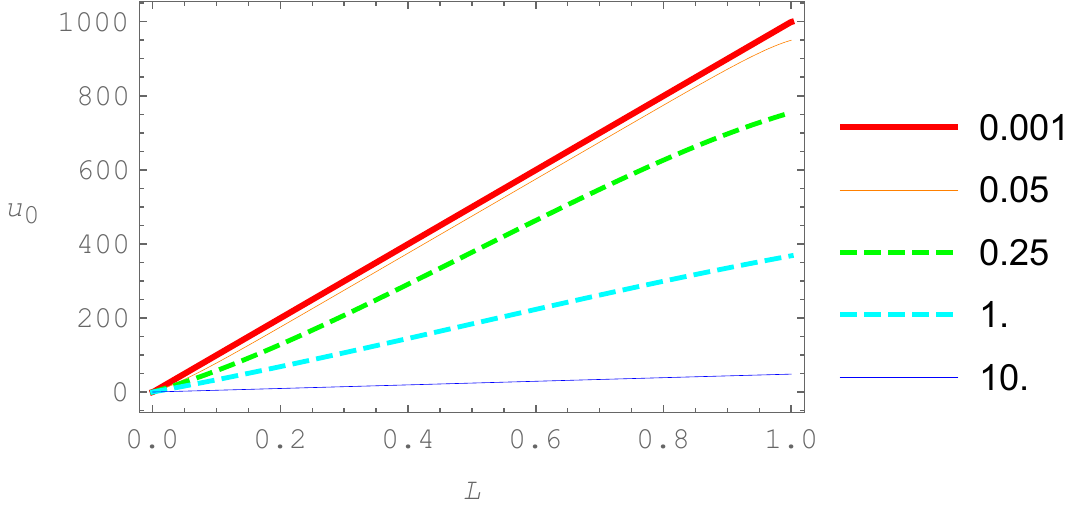}
	\caption{Distribution of $u_{0}$ along the beam loaded with the axial force and uniform temperature for various values of $\lambda$} 
	\label{fig:Ex2_1}
\end{figure}

Due to symmetricity of the cross section, first moment of the area $S_y$ vanishes. Thus, functions $\varPhi_\mathrm{N}(x)$ and $\varPhi_\mathrm{M}(x)$, Eqs.(\ref{eq:PhiNM}) are:
\begin{equation}\label{eq:Ex1_1}
\varPhi_\mathrm{N}(x)=  \frac{1}{A} \int _\Omega \alpha \Delta \theta \mathrm{d}A=\alpha \Delta \theta, \quad \varPhi_\mathrm{M}(x)=  \frac{1}{I_\mathrm{y}}\int _\Omega \alpha \Delta \theta z \mathrm{d}A=0.
\end{equation}
Since $\varPhi_\mathrm{M}(x)=0$, the transverse displacement $w$ equals zero for the whole domain. On the other side, the elastic (mechanical) axial force field is non-vanishing, homogeneous and equal to $N=P$. Equilibrium ensures that the internal force is not a function of axial coordinate, $N\neq f(x)$, so the axial displacement Eq.(\ref{eq:ODE}) is:
\begin{equation}\label{eq:Ex2_1}
u_{0}^{(1)}(x) = \alpha \Delta \theta+ \int_{0}^{L} \phi ({x-\xi)}  \frac{P}{E A}  \mathrm{d} \xi,
\end{equation}
where Eq.~(\ref{eq:Ex1_1})$_1$ was used. The solution follows from Eq.(\ref{eq:Solutionu0}) :
\begin{equation}\label{eq:Ex2_2}
 \frac{P}{E A}+\alpha \Delta \theta = -L_\lambda^{2} u_0^{(3)} + u_0^{(1)},
\end{equation}
where $(\alpha \Delta \theta) ^{(2)}=0$ due to constant heat expansion coefficient and temperature. This can be augmented with the constitutive boundary conditions Eq.~(\ref{eq:BCu0}):
\begin{equation}\label{eq:Ex2_3}
\begin{array}{l}
\displaystyle
u_0^{(2)}(0) -\frac{1}{L_\lambda}\left( u_0^{(1)}(0)-\alpha \Delta \theta \right) =0, 
\vspace{6pt}\cr
u_0^{(2)}(L) +\frac{1}{L_\lambda}\left( u_0^{(1)}(L)-\alpha \Delta \theta \right) =0.
\displaystyle
\end{array}
\end{equation}
Integration of the above problem yields:
\begin{equation}\label{eq:Ex2_4}
u_{0}(x)=\alpha \Delta \theta x + \frac{P x}{E A}- L_\lambda \frac{P}{2 E A } \left( e^{\frac{x}{L_\lambda}} -1 \right)  \left( e^{\frac{-x}{L_\lambda}}+e^{\frac{-L}{L_\lambda}}\right).
\end{equation}
The first and the second part represent the well-known local result while the third part accounts for the nonlocality. To provide a graphical insight into distribution of the axial displacement, values $P=1, L=1, EA=1, \alpha=0.1, \Delta \theta=0.1$ are chosen. Results for different values of $\lambda$ are given in Fig.~\ref{fig:Ex2_1}. A slight deviation from the linear distribution is noticeable at the beam's ends. The continuous reduction of the axial displacement under constant loading with the increase of the nonlocal parameter implies that increase of the nonlocal parameter effectively increases the beam stiffness.

The nonlocal strain is obtained by differentiation as:
\begin{equation}\label{eq:Ex2_5}
\varepsilon=\partial_x u_0= \alpha \Delta \theta + \frac{  P }{A E}-\frac{ P }{2 A E}  \left(e^\frac{x-L}{L_\lambda} + e^\frac{-x}{L_\lambda}\right).
\end{equation}

Straightforward application of the Remark 2 enables simple calculation of stress from the equilibrium equation as $\sigma=P/A$. So, although the obtained stress is equal to the local solution, the resulting strain and elongation are distinctly nonlocal.

Finally, the resulting axial displacement field could be solved by resorting to analysis of two decoupled problems. In particular, one can solve for thermal displacements in the first step, while the second step involves solution due to action of mechanical loading. The total solution is obtained as a linear combination of these results.

\subsection{Uniformly heated doubly clamped beam}
\label{Ex3}
In this example, a doubly clamped beam of similar geometrical and material properties like in the previous example is considered. The beam is clamped at both ends so it cannot expand. The total axial deformation is 
\begin{equation}\label{eq:Ex3_1}
u_0(x)=u_{0,\mathrm{th}}(x)+u_{0,\mathrm{el}}(x)=\alpha \Delta \theta x + u_{0,\mathrm{el}}(x),
\end{equation}
where the second - elastic part reflects the nonlocality caused by the reactions $P$ in supports. Also, the axial force does not change along the beam, so $N=P=\mathrm{const}$. Introducing nonlocality for the elastic part in the same manner as in Example \ref{Ex2} and exploiting Eq.~(\ref{eq:Ex3_1}), gives:
\begin{equation}\label{eq:Ex3_2}
u_0^{(1)}(x) -\alpha \Delta \theta = \int_{0}^{L} \phi ({x-\xi)}  \frac{P}{E A}  \mathrm{d} \xi .
\end{equation}
Considering Eq.(\ref{eq:Solutionu0}), the problem can be now stated as:
\begin{equation}\label{eq:Ex3_3}
\frac{P}{E A} + \alpha \Delta \theta = -L_\lambda^{2} (u_{0})^{(3)}+(u_{0})^{(1)},
\end{equation}
augmented with two boundary conditions:
\begin{equation}\label{eq:Ex3_4}
u_{0}^{(2)}(0)-\frac{1}{L_\lambda}(u_{0}^{(1)}(0)-\alpha \Delta \theta)=0, \qquad u_{0}^{(2)}(L)+\frac{1}{L_\lambda}(u_{0}^{(1)}(L)-\alpha \Delta \theta)=0.
\end{equation}
Additionally, since the beam is clamped it must be $u_0(0)=0$. The differential equation is of the third order, so only three boundary conditions can be accounted for at this point. The remaining one, $u_0(L)=0$, will be introduced at the later stage. Because this problem has the identical structure as the one in Example \ref{Ex2}, resulting displacement distribution is identical, cf. Eq.(\ref{eq:Ex2_4}):
\begin{equation}\label{eq:Ex3_6}
u_0(x)=\alpha \Delta \theta x+\frac{ P}{EA }x- L_\lambda \frac{ P }{2 EA }    \left(e^{\frac{x}{L_\lambda}}-1\right) \left(e^{\frac{-L}{L_\lambda}}+e^{\frac{-x}{L_\lambda}}\right),
\end{equation}
with the difference that this equation contains unknown constant - reaction $P$. The reaction can be determined by exploiting condition $u_0(L)=0$ in Eq.~(\ref{eq:Ex3_6}), giving:
\begin{equation}\label{eq:Ex3_7}
P = -\frac{\alpha \Delta \theta A E L  e^{L/ L_\lambda}}{e^{L/ L_\lambda} (L- L_\lambda)+ L_\lambda}.
\end{equation}
Introducing the reaction $P$ into Eq.~(\ref{eq:Ex3_6}) gives the final distribution of the axial displacement:
\begin{equation}\label{eq:Ex3_8}
u_0(x)=-\frac{\alpha \Delta \theta  L_\lambda \left(-L e^{\frac{x}{ L_\lambda}}+\left(e^{\frac{L}{ L_\lambda}}-1\right) (2 x-L) +L e^{\frac{L-x}{L_\lambda}} \right)}{2(e^{\frac{L}{L_\lambda}} (L- L_\lambda)+ L_\lambda)}.
\end{equation}
Contrary to the statically determine Example \ref{Ex2}, this solution involves coupled thermal and elastic part that cannot be separated into two decoupled problems in the manner suggested in the former example.

\begin{figure}
	\centering
	\includegraphics[scale=0.45]{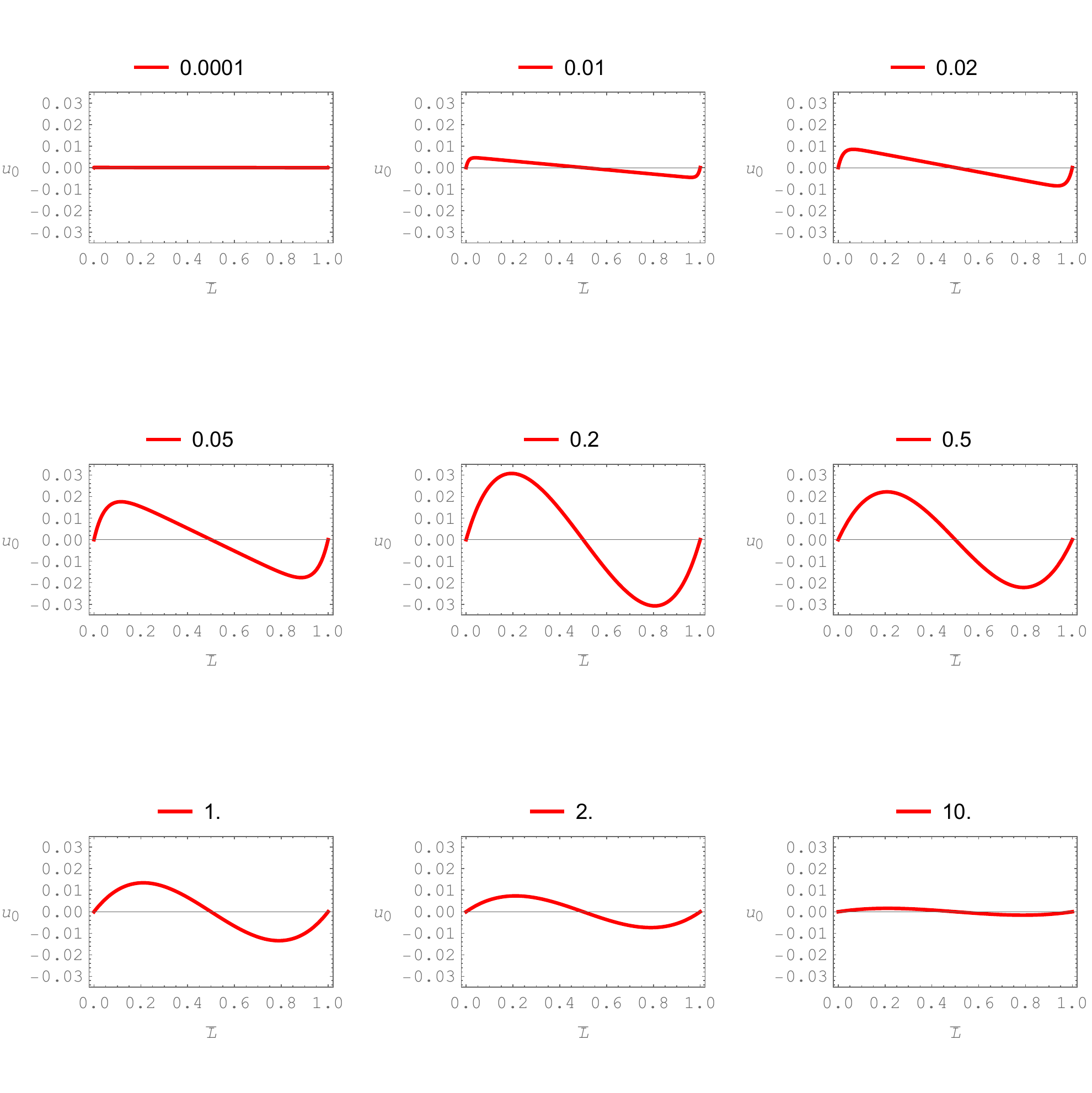}
	\caption{Distribution of $u_0$ along the doubly clamped beam loaded with uniform temperature for various values of $\lambda$} 
	\label{fig:Ex3_1}
\end{figure}

\begin{figure}
	\centering
	\includegraphics[scale=0.40]{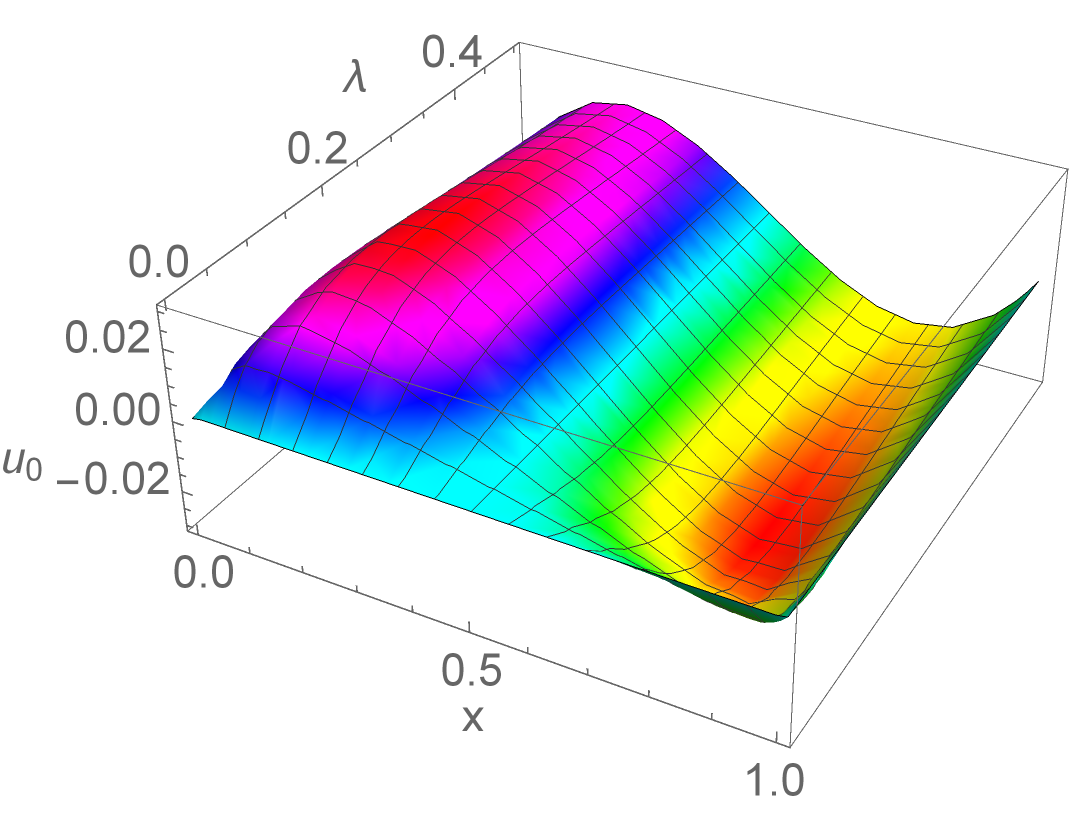}
	\caption{A three-dimensional detail of distribution of $u_0$ along the doubly clamped beam loaded with uniform temperature for $\lambda \in \left\langle 0,0.4\right] $ } 
	\label{fig:Ex3_3}
\end{figure}

To present results graphically, the same parameters as in Example \ref{Ex2} were used. Obtained total axial displacements $u_0$ are given in Fig.~\ref{fig:Ex3_1} for a set of nonlocal parameters. Since the local formulation will lead toward homogeneous solution $u_0(x)=0$ (simulated here with $\lambda=0.0001$), the effect of nonlocality is evident through the non-homogeneous displacement field. Inspection of Fig.~\ref{fig:Ex3_1} reveals that the nanobeam appears to initially becomes more flexible manifested by larger axial displacements, but after some threshold value of the nonlocal parameter, it becomes more rigid. It is also interesting to note that the nonlocal solution for the increasing value of the nonlocal parameter ($\lambda=10$) converge to the local solution ($\lambda=0$). The three-dimensional representation, Fig.~\ref{fig:Ex3_3}, shows that the lowest turning point is slightly below $\lambda=0.2$.  The starting softening behaviour is due to the fact that the local axial displacements are vanishing, so that for small values of $\lambda$ the nonlocal axial displacements are different from zero but as $\lambda$ tends to infinity the stress-driven total deformation goes to zero. A starting soft behaviour must be associated with any nonlocal model (strain-driven or stress-driven
formulations) since local displacement (in this example) are vanishing. Similar behaviour was previously noticed in \cite{Romano17b} in the bending of a beam. An important conclusion observed from the present example is that the terms "stiffer" and "softer" (widely used in literature) with respect to the nonlocal parameter are often misleading and often meaningless. This observation is motivated from the fact that the nonlocal total deformation is a field, so that the sign changes are often unpredictable.

\begin{figure}
	\centering
	\includegraphics[scale=0.45]{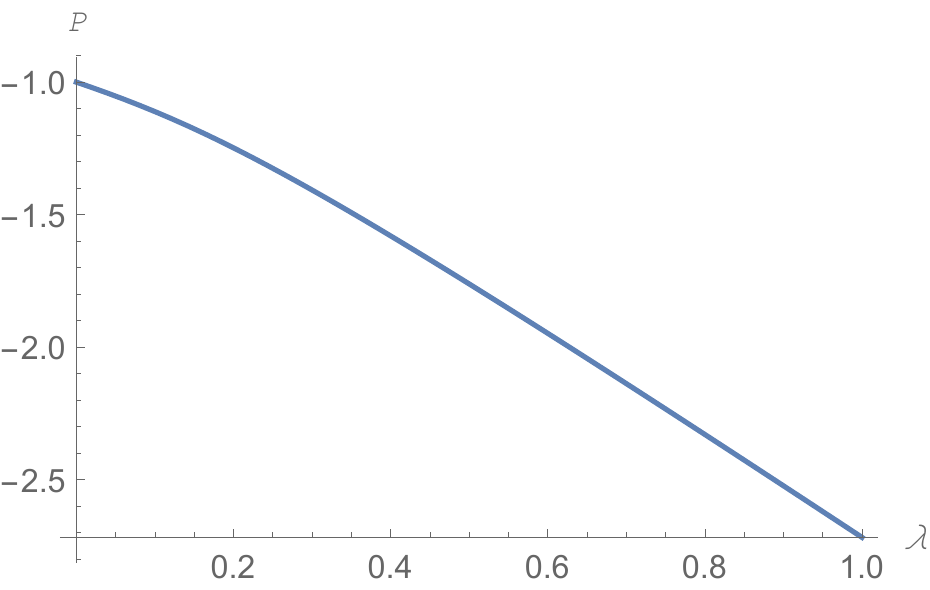}
	\caption{Dependence of the reaction in supports $P$ on the nonloal parameter $\lambda$.}  
	\label{fig:Ex3_4}
\end{figure}

Results presented in the previous example showed that increase in the nonlocal parameter will increase the nanobeam stiffness. However, in this example, the nonlocal elastic shortening at the nanobeam ends must match the thermal elongation that has the local character. This means the support reactions cannot remain constant but have to increase as well in order to compensate for increased stiffness due to $\lambda$. Fig.~\ref{fig:Ex3_4} clearly illustrates this dependency on the nonlocal parameter. Therefore, due to presence of additional kinematical constraints that relate nonlocal kinematical quantity (mechanical shortening due to support reactions) to a local kinematical quantity (thermal elongation) in statically indeterminate problems, one inevitably obtains reactions that depend on the constitutive behaviour. The same constitutive dependence can be observed even in the local case, where support reactions depend on the coefficient of heat expansion.

\begin{figure}
	\centering
	\includegraphics[scale=0.40]{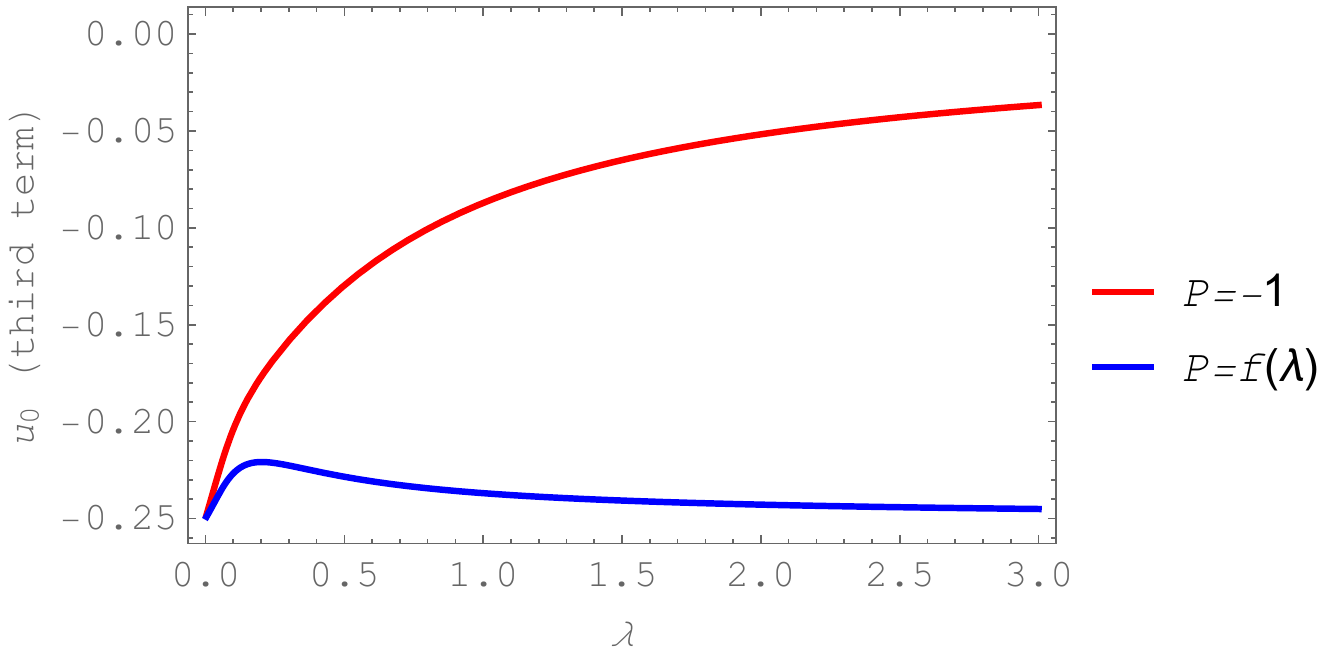}
	\caption{Dependence of the third term in Eq.~(\ref{eq:Ex3_6}) on $\lambda$ for a constant axial force and the one where $P$ is dependent on the nonlocal parameter.}  
	\label{fig:Ex3_5}
\end{figure}

 Note that curve $P$ vs. $\lambda$ is initially slightly curved and later almost linear,  Fig.~\ref{fig:Ex3_4}. This can be explained by a careful analysis of Eq.~(\ref{eq:Ex3_6}). In particular, the first and the second term are linear, while the third is nonlinear. Hence, the third term is responsible for reduction/increase of stiffness. For the smaller values of $\lambda$, the third term has some influence, visible as the initial nonlinearity in Fig.~\ref{fig:Ex3_4}. As $\lambda$ increases, exponential terms change only slightly so the third term becomes almost a constant. As a consequence, the sum of the second and the third term tends to be a linear form for larger $\lambda$. A comparison of behaviour of the third term for a constant value of $P$ (i.e. Example \ref{Ex2}) and for a variable one from the present example is provided in Fig.~\ref{fig:Ex3_5}. Such behaviour was not noticed in \cite{Barretta17_2} in the isothermal statically indeterimine problems, where monotonic increase in the nanobeam stiffness with the increase in the nonlocal parameter was observed. The principal difference between these two cases is the fact that in the isothermal statically indetermine problems only prescribed displacement in a point (usually equal to zero) is given. In the present case, a part of solution (thermal elongation) is known in advance for the whole domain, so this could be considered as an additional constraint on the allowable forms of axial displacement.

\begin{figure}
	\centering
	\includegraphics[scale=0.50]{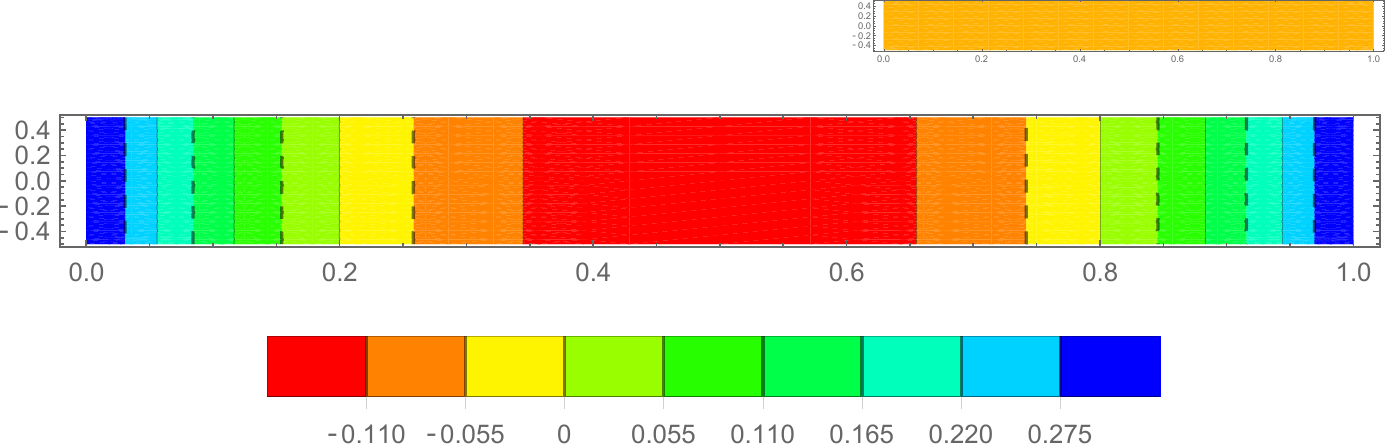}
	\caption{Distribution of the normal strain $\varepsilon$ along the doubly clamped beam loaded with uniform temperature, $\lambda=0.1$, lateral view. Inset in the top right corner - normal stress distribution $\sigma=-1.33$ used to obtain the normal strain.}  
	\label{fig:Ex3_2}
\end{figure}

The normal strain $\varepsilon(x)$ distribution is given in Fig.~\ref{fig:Ex3_2}.  Since the local formulation results in the homogeneous normal strain field, in the present case nonlocal behaviour for $\lambda=0.1$ is clearly visible once more. Inset in the top right corner presents the homogeneous stress field as obtained from this formulation.

\subsection{Bending of a nonlocal beam due to nonuniform temperature field in the transverse direction}
In this example, the identical problem as in \cite{Canadija16a} in considered, thus providing a basis for comparison of the gradient and the integral approach. The cantilever beam's length is $L$, while the rectangular cross-section is defined by the height $h$ and the width $b$. The temperature field is given as $\theta(z)=\theta_0+a_1 e^{a_2(z+\frac{h}{2})}$, where $a_1$ and $a_2$ are constants defining distribution of temperature. The beam is not loaded by a mechanical load in any way. Thus, due to its nonlinear temperature dependency on the transverse coordinate, the nanobeam will exhibit in-plane bending. It will be assumed that the initial temperature was homogeneous and equal to zero, $\theta_0=0$.

Functions $\varPhi_\mathrm{N}$ and $\varPhi_\mathrm{M}$, Eq.~(\ref{eq:PhiNM}), are obtained from the provided temperature field as:
\begin{equation}\label{eq:Ex4_1}
\begin{array}{l}
\varPhi_\mathrm{N}= \frac{1}{A} \int_{\Omega} \alpha \Delta \theta \mathrm{d}A=\frac{\alpha  a_1  \left(e^{a_2 h}-1\right)}{ a_2 h},\\
\varPhi_\mathrm{M}= \frac{1}{I_\mathrm{y}} \int_{\Omega} \alpha \Delta \theta z \mathrm{d}A=\frac{6 \alpha a_1 \left(a_2 h+e^{a_2h} (a_2 h-2)+2\right)}{ {a_2}^2 h^3}.
\end{array}
\end{equation}
Since the temperature field does not depend on the longitudinal coordinate, both functions are reduced to constants and their derivatives will vanish. Also, due to absence of external mechanical loads $M=0$. The transverse displacement is the solution of differential equation Eq.~(\ref{eq:v}):
\begin{equation}\label{eq:Ex4_2}
0=  L_\lambda^{2} w^{(4)} - w^{(2)}-\varPhi_\mathrm{M},
\end{equation}
augmented with the boundary conditions, Eq.~(\ref{eq:BCv}):
\begin{equation}\label{eq:Ex4_3}
\begin{array}{c}
-w^{(3)}(0) -\frac{1}{L_\lambda} \left( -w^{(2)}(0)-\varPhi_\mathrm{M} \right) =0, \\
-w^{(3)}(L) + \frac{1}{L_\lambda} \left( -w^{(2)}(L)-\varPhi_\mathrm{M} \right) =0, \\
w(0) =0,\qquad w'(0)=-\varphi(0)=0.
\end{array}
\end{equation}
The solution of the above problem is:
\begin{equation}\label{eq:Ex4_5}
\begin{array}{l}
w(x)=-\frac{ 3 \alpha  a_1 \left(e^{a_2 h} (a_2 h-2)+a_2 h+2\right)}{{a_2}^2 h^3} x^2 ,
\end{array}
\end{equation}
what is identical to the solution based on the gradient formulation, cf. \cite{Canadija16a}. The same conclusion can be stated for the rotation:
\begin{equation}\label{eq:Ex4_6}
\varphi(x)=-\partial_x w(x)= \frac{ 6 \alpha  a_1 \left(e^{a_2 h} (a_2 h-2)+a_2 h+2\right)}{{a_2}^2 h^3} x.
\end{equation}
The average axial displacement follows from the differential equation, Eq.~(\ref{eq:Solutionu0}):
\begin{equation}\label{eq:Ex4_7}
\varPhi_\mathrm{N} = -L_\lambda^{2} u_\mathrm{0}^{(3)}+u_\mathrm{0}^{(1)},
\end{equation}
with the boundary conditions:
\begin{equation}\label{eq:Ex4_8}
\begin{array}{c}
u_\mathrm{0}^{(2)}(0)-\frac{1}{L_\lambda}(u_\mathrm{0}^{(1)}(0)-\varPhi_\mathrm{N})=0, \qquad u_\mathrm{0}^{(2)}(L)+\frac{1}{L_\lambda}(u_\mathrm{0}^{(1)}(L)-\varPhi_\mathrm{N})=0, \qquad u_\mathrm{0}(0) =0.
\end{array}
\end{equation}
The solution is:
\begin{equation}\label{eq:Ex4_9}
u_\mathrm{0}=\frac{ a_1 \alpha  x \left( e^{a_2 h}-1 \right)}{a_2 h} ,
\end{equation}
what again corresponds to the gradient solution. Obviously, the integral formulation does not show nonlocal behaviour as well, due to particular temperature distribution and absence of external mechanical loading.

Stress can be conveniently determined by application of Eq.~(\ref{eq:ThStress2}). Since the temperature field is independent of the longitudinal coordinate, the expression is reduced to:
\begin{equation}\label{eq:Ex4_10}
\sigma(x,z) = E \left\lbrace -\alpha \Delta \theta(x,z)+\varPhi_\mathrm{N}(x)-z \; \varPhi_\mathrm{M}(x) \right\rbrace .
\end{equation}
This finally yields:
\begin{equation}\label{eq:Ex4_11}
\sigma(z) = \alpha a_1 E \left\lbrace - e^{a_2(z+\frac{h}{2})}+\frac{e^{a_2 h}-1}{a_2 h}-\frac{6 z \left(a_2 h+e^{a_2 h} (a_2 h-2)+2\right)}{a_2^2 h^3} \right\rbrace .
\end{equation}
As expected, the stresses are not affected by nonlocality and should be obtained from the local form, Eq.~(\ref{eq:IntForcesMT}) as well. The thermal axial force and the thermal bending moment are evaluated as:
\begin{equation}\label{eq:Ex4_12}
\begin{array}{c}
N_\mathrm{th}=\int_{\Omega} \alpha \Delta \theta E\mathrm{d}A= \frac{\alpha  a_1 b E \left(e^{a_2 h}-1\right)}{a_2},\\
M_\mathrm{th}=\int_{\Omega} \alpha E \Delta \theta z \mathrm{d}A=\frac{\alpha a_1 b E \left(a_2 h+e^{a_2 h} (a_2 h-2)+2\right)}{2 a_2^2}.
\end{array}
\end{equation}
The above results introduced in Eq.~(\ref{eq:IntForcesMT}) provides desired verification. Note that stress dependency on the nonlocal parameter could be obtained if $\partial _x ^2 \varPhi_\mathrm{N}$ and $\partial _x ^2 \varPhi_\mathrm{M}$ does not vanish, Eq.~(\ref{eq:ThStress2}).

\subsection{Nonuniform heating of a doubly clamped beam}
\label{Ex4}
In this example a doubly clamped beam $L=100$ nm long is considered. The cross-section is situated in $y-z$ plane and is assumed to be of a square shape with height $h=1$ nm and width $b=1$ nm. The Young's modulus is $E=1$ nN/nm$^2$. The coefficient of thermal expansion is $\alpha=1$ 1/K. The temperature increase varies along the beam as:
\begin{equation}\label{eq:Ex5_1}
\theta (x,z)=\frac{{a_0} x^2 \left(\frac{h}{2}+z\right)^2}{h^2},
\end{equation}
where $a_0=1$ K/nm$^2$. Consequently, temperature is a function of both  longitudinal coordinate $x$ and the transverse coordinate $z$. Note that the plane $z=-h/2$ does not expand due to temperature variation. As a consequence, $\varPhi_\mathrm{N}$ and $\varPhi_\mathrm{M}$ are now functions of $x$:
\begin{equation}\label{eq:Ex5_1a}
\varPhi_\mathrm{N}(x)= \frac{1}{3} a_0 \alpha x^2,   \quad   \varPhi_\mathrm{M}(x)=\frac{1}{h} a_0 \alpha x^2,
\end{equation}
and their first and second derivative do not vanish. At the coordinate system's origin $x=0$, both functions and their derivatives take null values.

The bending moment and the axial force are distributed as follows:
\begin{equation}\label{eq:Ex5_2}
M=M_\mathrm{A}+P_{\mathrm{Az}}x,\qquad N=P_\mathrm{Ax},
\end{equation}
where $M_\mathrm{A}, P_{\mathrm{Ax}}, P_\mathrm{Az}$ are unknown bending moment and reaction forces in the support A. Equilibrium constraints provide additional equations for determination of the reactions at the support B. Initially, distribution of the transverse displacements will be determined from the differential equation Eq.~(\ref{eq:v}):
\begin{equation}\label{eq:Ex5_3}
\frac{M}{E I_\mathrm{y}} = L_\lambda^{2} \left\lbrace  w^{(4)}+\varPhi_\mathrm{M}^{(2)} \right\rbrace  -\left( w^{(2)}+\varPhi_\mathrm{M}\right) ,
\end{equation}
augmented with the boundary conditions:
\begin{equation}\label{eq:Ex5_4}
\begin{array}{c}
w^{(3)}(0) -\frac{1}{L_\lambda} w^{(2)}(0)=0, \\
w^{(3)}(L)+\varPhi_\mathrm{M}^{(1)}(L) +\frac{1}{L_\lambda} \left( w^{(2)}(L)+\varPhi_\mathrm{M}(L) \right) =0. \\
w(0) =0,\qquad w'(0)=-\varphi(0)=0.
\end{array}
\end{equation}

Since the differential equation is of the fourth order, only four boundary conditions can be specified at this point. This will provide a solution that involves unknown reactions in support A, $M_\mathrm{A}$ and $P_\mathrm{Az}$. To determine unknown reactions, the remaining boundary conditions in the support B, $v(L) =0, v'(L)=-\varphi(L)=0$, have to be exploited. Determination of these can be also suitably performed by Wolfram Mathematica, but due to lengthy forms, support reactions and deflection line equations are not reported. It is just pointed out that reactions show dependence on the nonlocal parameter in the line with the discussion in Example \ref{Ex3}.

As for the axial displacements, these can be determined from Eq.~(\ref{eq:Solutionu0}) as:
\begin{equation}\label{eq:Ex5_8}
\frac{N}{E A}  = -L_\lambda^{2} \left\lbrace  u_0^{(3)}- \varPhi_\mathrm{N} ^{(2)} \right\rbrace + u_0^{(1)}-\varPhi_\mathrm{N},
\end{equation}
together with the boundary conditions:
\begin{equation}\label{eq:Ex5_9}
\begin{array}{l}
u_0^{(2)}(0) -\frac{1}{L_\lambda}  u_0^{(1)}(0)  =0, \\
u_0^{(2)}(L)-\varPhi_\mathrm{N}^{(1)}(L) +\frac{1}{L_\lambda}\left( u_0^{(1)}(L)-\varPhi_\mathrm{N}(L) \right) =0, \\
u_0 (0)=0.
\end{array}
\end{equation}
The solution includes unknown axial reaction $P_\mathrm{Ax}$ at the support A. This reaction follows from $u_0 (L) =0$. 

\begin{figure}
	\centering
	\includegraphics[scale=0.45]{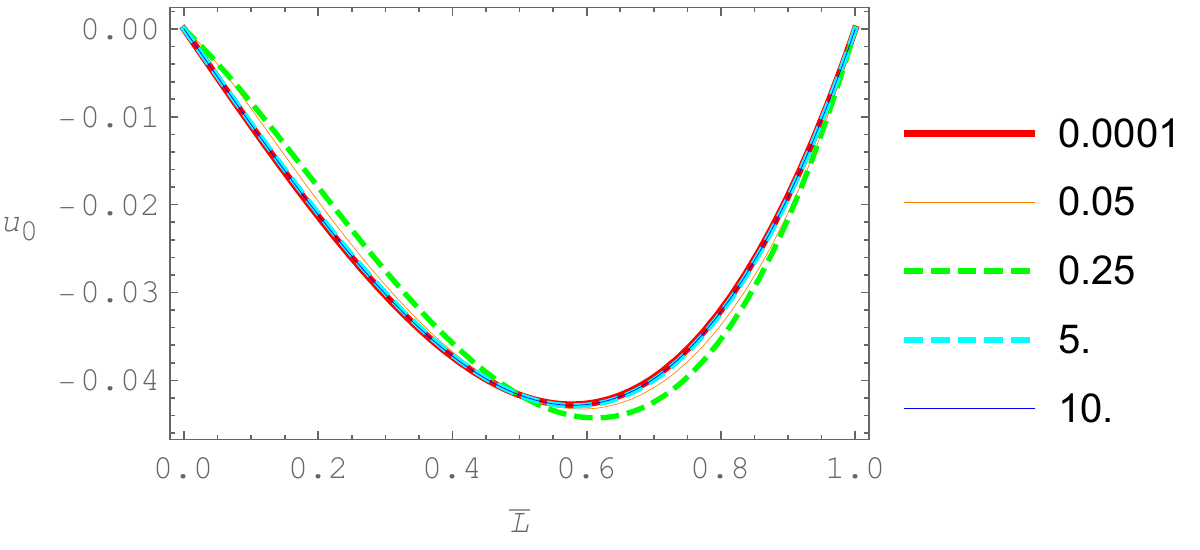}
	\caption{Distribution of the axial displacement $u_0$ along the doubly clamped beam loaded with the non-uniform temperature for $\lambda \in \left\lbrace 10^{-4}, 0.05, 0.25, 5, 10\right\rbrace $ nm } 
	\label{fig:Ex5_1}
\end{figure}

\begin{figure}
	\centering
	\includegraphics[scale=0.45]{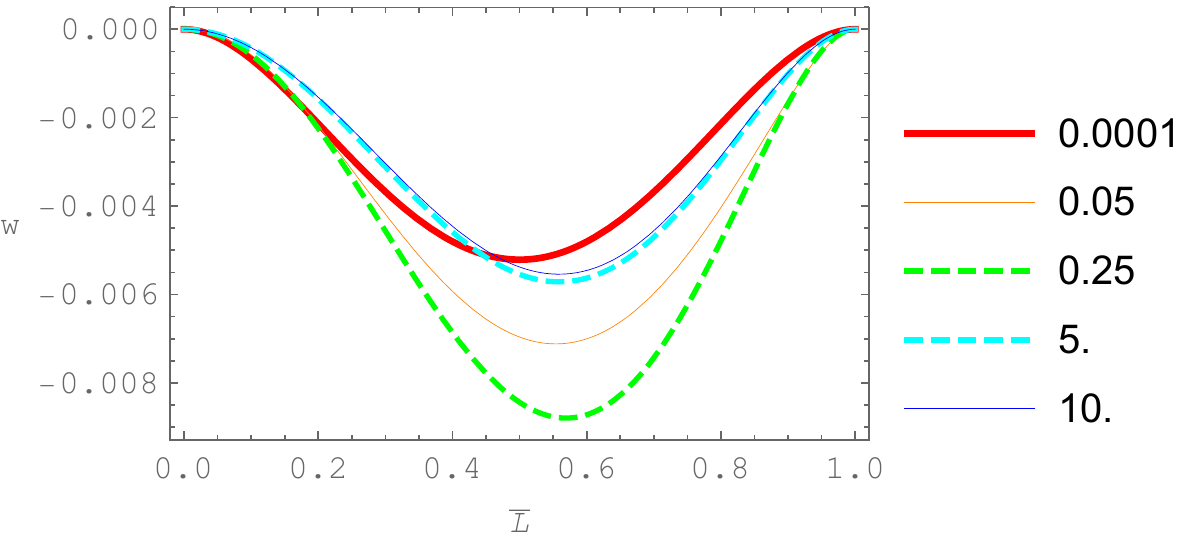}
	\caption{Distribution of the transverse displacement $v$ along the doubly clamped beam loaded with the non-uniform temperature for $\lambda \in \left\lbrace 10^{-4}, 0.05, 0.25, 5, 10\right\rbrace $ nm } 
	\label{fig:Ex5_2}
\end{figure}

\begin{figure}
\centering
\includegraphics[scale=0.45]{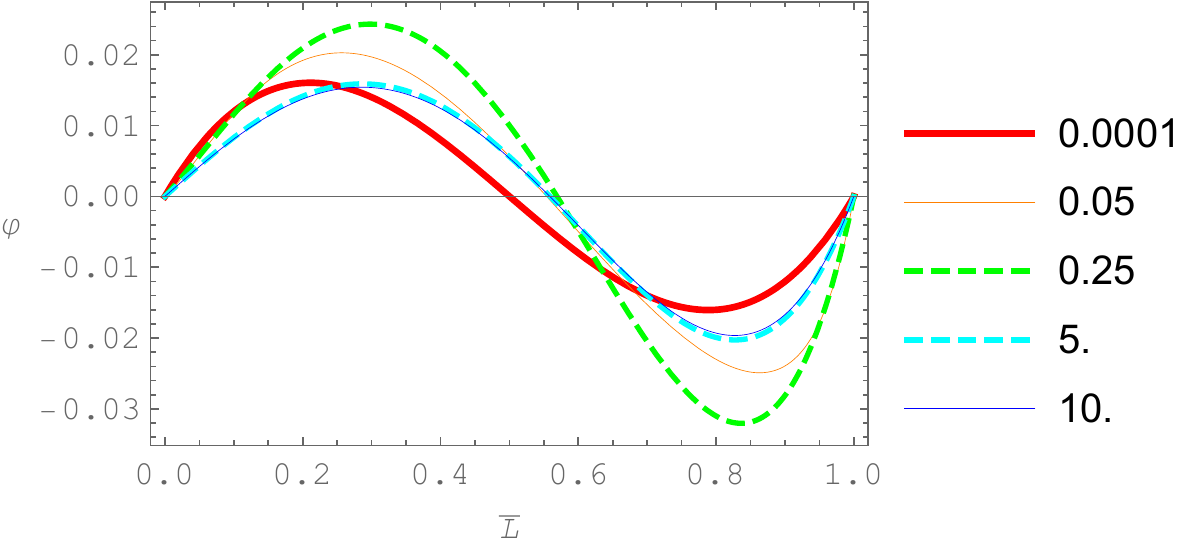}
\caption{Distribution of the bending angle $\varphi$ along the doubly clamped beam loaded with the non-uniform temperature for $\lambda \in \left\lbrace 10^{-4}, 0.05, 0.25, 5, 10\right\rbrace $ nm } 
\label{fig:Ex5_3}
\end{figure}

\begin{figure}
	\centering
	\includegraphics[scale=0.45]{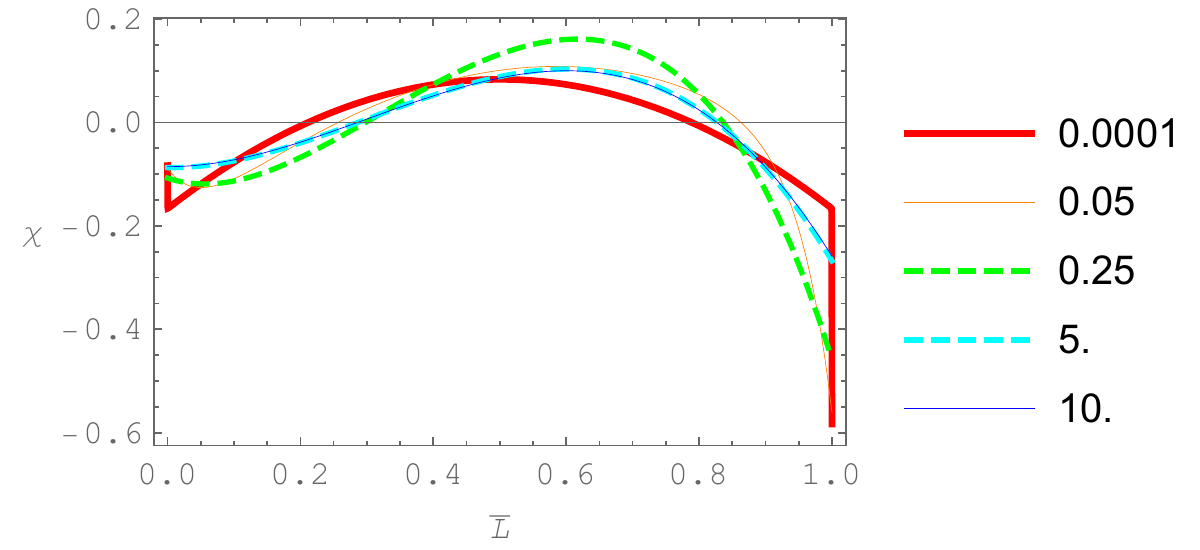}
	\caption{Distribution of the curvature $\chi$ along the doubly clamped beam loaded with the non-uniform temperature for $\lambda \in \left\lbrace 10^{-4}, 0.05, 0.25,  5, 10\right\rbrace $ nm } 
	\label{fig:Ex5_3b}
\end{figure}

\begin{figure}
	\centering
	\includegraphics[scale=0.45]{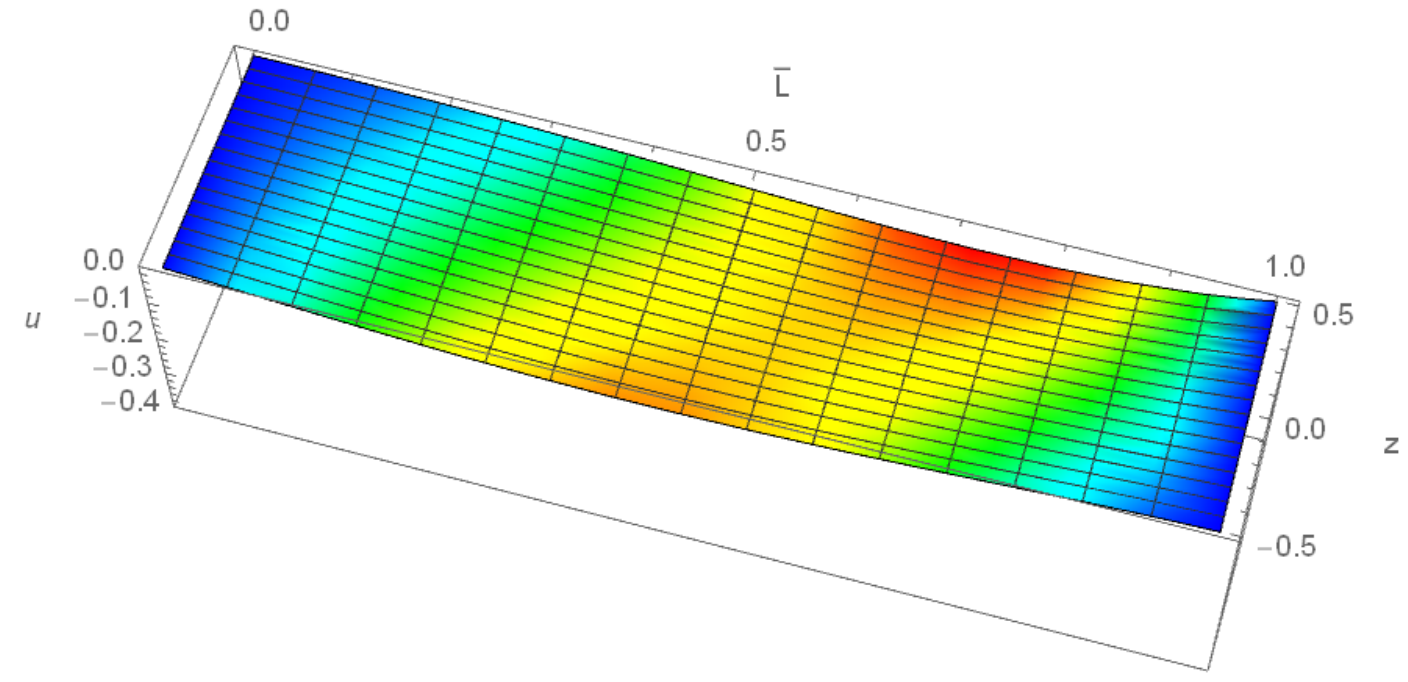}
	\caption{Distribution of the axial displacement $u(x,z)$ along the doubly clamped beam loaded with the non-uniform temperature, lateral view.  $\lambda=0.25$ nm} 
	\label{fig:Ex5_4}
\end{figure}

\begin{figure}
	\centering
	\includegraphics[scale=0.4]{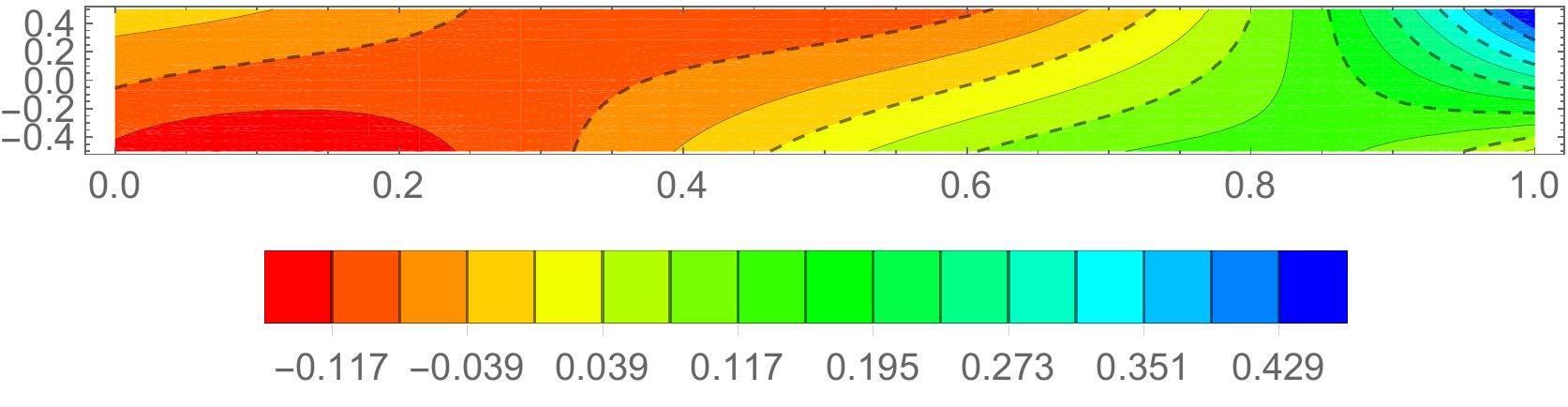}
	\caption{Distribution of the normal strain $\varepsilon=\varepsilon_{\mathrm{th}}+\varepsilon_{\mathrm{el}}$ along the doubly clamped beam loaded with the non-uniform temperature, lateral view. $\lambda=0.25$ nm } 
	\label{fig:Ex5_5}
\end{figure}

The main results are summarized in Figs.~(\ref{fig:Ex5_1}-\ref{fig:Ex5_5}). To make figures more easier to interpret, the length was normalized and denoted with an overbar. The minimal value for $\lambda$ serves as the limit case that corresponds to the classical Bernoulli-Euler local solution. Similarly to Example \ref{Ex3}, the nanobeam behaviour tends to be more flexible until a certain threshold is reached and subsequently converge to another, more rigid curve.  

The solution for the axial displacement $u(x,z)$ is graphically presented in Fig.~\ref{fig:Ex5_4} for a selected value of the nonlocal parameter. Provided distribution shows that boundary conditions are fulfilled at the left and right support. The maximal longitudinal displacement occurs at approximately 60\% of the beam's length.

\begin{figure}
	\centering
	\includegraphics[scale=0.4]{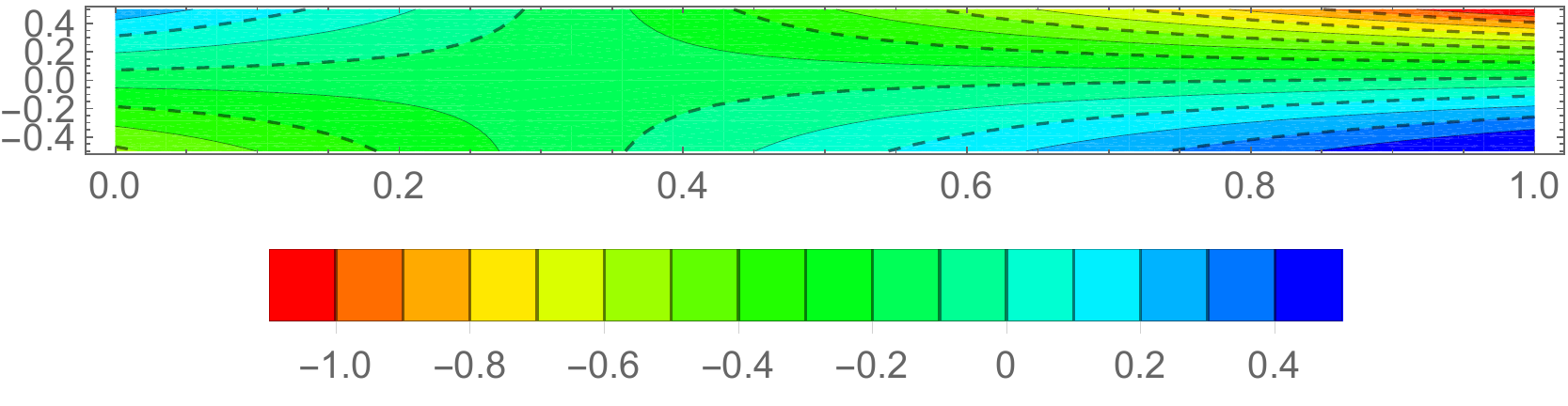}
	\caption{Distribution of the normal stress $\sigma$ along the doubly clamped beam loaded with the non-uniform temperature, lateral view. $\lambda=0.25$ nm } 
	\label{fig:Ex5_6}
\end{figure}

Finally, Fig.~\ref{fig:Ex5_5} provides an insight into distribution of the nonlocal strain $\varepsilon(x,z)$. Compared to the more simpler case in Example \ref{Ex3}, Fig.~\ref{fig:Ex3_2}, the normal strain shows significantly more complex pattern. The accompanying stress field is presented in Fig.~\ref{fig:Ex5_6}. Note that the local solution for stresses would provide the same distribution as for the normal strain with a difference being only due to multiplication by Young's modulus $E$. Difference in the distribution pattern in Fig.~\ref{fig:Ex5_5} and Fig.~\ref{fig:Ex5_6} clearly indicates the nonlocal behaviour.


\break

\section{Conclusions}
\label{Conclusions}
A new thermoelastic integral model for nanobeams has been developed
by adopting the nonlocal elasticity theory by G. Romano and R. Barretta \cite{Romano2017c}. 
Effectiveness of the presented methodology has been tested by examining selected case-studies.

Some closing remarks are collected below.

\begin{enumerate}
	\item 
	The new stress-driven integral thermoelastic law, equipped with the averaging kernel introduced by Helmholtz, is equivalent to a set of differential equations and to suitable higher-order constitutive boundary conditions.
	\item 
	The example section also demonstrates that thermoelastic nonlocal solutions do not necessarily lead toward gradually stiffer solution for increasing values of the nonlocal parameter. 
	Such behaviour has been displayed in statically indeterminate structures. 
	Otherwise, solutions show monotonically increasing stiffness like in isothermal cases.
	\item 
	Since the present method is stress-driven, the dependence of the boundary conditions on the constitutive behaviour is circumvented in the statically determine problems, as demonstrated in earlier researches within isothermal framework. As discussed in Example \ref{Ex3} and further confirmed in Example \ref{Ex4}, in the statically indeterminate cases the reactions still show dependence on the nonlocal parameter due to coupling of the local (thermal expansion/contraction) and nonlocal (mechanical) terms. 
	\item Having latter remark in mind, proposed formulation can be considered as the mixed local-nonlocal approach.
\end{enumerate}

\section{Acknowledgements}
\label{Acknowledgements}
This work has been supported in part by Croatian Science Foundation under the project no. 6876 - Assessment of structural behaviour in limit state operating conditions and in part  under the project ReLUIS - Italian Department of the Civil Protection 2017. This support is gratefully acknowledged.

\bigskip

\noindent

{\bf References}

\bibliographystyle{elsarticle-harv}
\bibliography{SmallSizeParameter}

\end{document}